\documentclass[conference]{IEEEtran}
\pdfoutput=1
\IEEEoverridecommandlockouts
\usepackage{multirow}
\usepackage{amsmath,amssymb,amsfonts}
\usepackage{algorithmic}
\usepackage{graphicx}
\usepackage{textcomp}
\usepackage{xcolor}
\usepackage{soul}
\usepackage{microtype}

\usepackage[style=ieee]{biblatex}
\usepackage[bookmarks=false,breaklinks=false,pdfborder={0 0 1},backref=false,colorlinks=false]{hyperref}
\hypersetup{pdftitle={Towards a Fault-Injection Benchmarking Suite},
 pdfauthor={Tianhao Wang, Robin Thunig, Horst Schirmeier},
 backref = true, pagebackref = true, colorlinks = true, linkcolor = black, anchorcolor = black, citecolor = black, filecolor = black, urlcolor = black, pagecolor= black}

\usepackage{prettyref}
\newrefformat{cha}{\hyperref[#1]{Chap.~\ref*{#1}}}
\newrefformat{sec}{\hyperref[#1]{Sec.~\ref*{#1}}}
\newrefformat{sub}{\hyperref[#1]{Sec.~\ref*{#1}}}
\newrefformat{subsec}{\hyperref[#1]{Sec.~\ref*{#1}}}
\newrefformat{tab}{\hyperref[#1]{Tab.~\ref*{#1}}}
\newrefformat{fig}{\hyperref[#1]{Fig.~\ref*{#1}}}
\newrefformat{lst}{\hyperref[#1]{List.~\ref*{#1}}}
\newrefformat{pat}{\hyperref[#1]{Patch~\ref*{#1}}}
\newrefformat{alg}{\hyperref[#1]{Alg.~\ref*{#1}}}
\newrefformat{eq}{\hyperref[#1]{Eq.~\ref*{#1}}}

\begin{document}
\title{Towards a Fault-Injection Benchmarking Suite}

\author{\IEEEauthorblockN{Tianhao Wang, Robin Thunig and Horst Schirmeier}
	\IEEEauthorblockA{Chair of Operating Systems, TU Dresden, Germany \\
		tianhao.wang2@mailbox.tu-dresden.de\\
		\{robin.thunig, horst.schirmeier\}@tu-dresden.de
	}
}

\maketitle

\begin{abstract}
    Soft errors in memories and logic circuits are known to disturb program
    execution. In this context, the research community has been proposing a
    plethora of fault-tolerance (FT) solutions over the last decades, as well as
    fault-injection (FI) approaches to test, measure and compare them. However,
    there is no agreed-upon benchmarking suite for demonstrating FT or FI
    approaches. As a replacement, authors pick benchmarks from other domains,
    e.g. embedded systems. This leads to little comparability across
    publications, and causes behavioral overlap within benchmarks that were not
    selected for orthogonality in the FT\slash{}FI domain.

    In this paper, we want to initiate a discussion on what a benchmarking suite
    for the FT\slash{}FI domain should look like, and propose criteria for
    benchmark selection.
\end{abstract}

\section{Introduction}
\label{sec:intro}

Soft errors have been posing threats to modern computer systems, particularly in
mission-critical applications such as satellites. Software-implemented
hardware fault
tolerance is a cost-effective way of mitigating soft errors.
Researchers typically demonstrate their fault-tolerance (FT) mechanisms
or fault-injection (FI) frameworks with a set of benchmarking
programs. While there exists abundant research on FT and FI, to the best of our
knowledge, there is no dedicated benchmarking suite for the FT\slash{}FI
domain.

To test their FT and FI mechanisms, researchers usually resort to
benchmarking suites from other domains such as
\textit{TACLeBench} \cite{TACLeBench} for \textit{Worst-Case Execution Time (WCET)} research
or \textit{MiBench}
\cite{guthaus:2001:mibench} for embedded systems. These benchmarking suites are
designed for different purposes and metrics than those relevant in the FT\slash{}FI
domain. This leads to three major shortcomings:

(1) \textit{Little Comparability across FT\slash{}FI Papers.}
It is hard to compare the effectiveness of different FT\slash FI mechanisms when
they are tested with different benchmarking programs. We also show in
\prettyref{sec:evaluation} that even the same benchmarking program could have a
dramatically different fault space when built with different runtime.

(2) \textit{Overlapping Benchmarks}.
It is difficult to select a minimal set of benchmarks from other domains that
achieves a full coverage of FT\slash{}FI relevant properties. Researchers
usually pick as many benchmarks as possible to deliver a convincing
demonstration, regardless of the potential overlap within the benchmarks. This
leads to inefficiency as each additional benchmark could cost a huge amount of
computing resources and time, particularly in the FI experiments.

(3) \textit{Limited Configurability}.
Technically, the out-of-the-box benchmarks from other domains are usually not
optimal for FI campaigns. For example, we often need programs
to meet certain requirements such as a minimal amount of
memory-access events or a maximal execution time. However, in most cases
-- if at all --
we can only adjust the scale of the input.

Therefore we would like to open a discussion whether the FT\slash FI domain
needs its own benchmarking suite, and which requirements it should meet.

\section{Preferable Benchmark Properties}
\label{sec:main}

As an example, Guthaus et al.~\cite{guthaus:2001:mibench} select benchmarks for
\textit{MiBench} on a basis
of their relevance in industrial embedded systems, and primarily focus on the
system performance as a benchmarking metric.
Falk et al.'s \textit{TACLeBench}
\cite{TACLeBench} is apt for WCET research.
The focus is on the metrics of WCET analysis tools.
The authors further differentiate usage-type categories like
\textit{kernel},
\textit{sequential},
\textit{application},
\textit{test} and
\textit{parallel}.

While these domain-specific criteria and metrics could also be important for
FT\slash{}FI research, they provide us with very limited insight when analysing
the fault tolerance of programs. Some benchmarks could also be totally undesired
by FT\slash FI experiments, such as the \textit{test} group in
\textit{TACLeBench} or \textit{MiBench}'s \textit{basicmath} benchmark, which are
designed to stress test the system under test with a repeated and simple workload.

From the experience and observations in the literature, we would like to propose several
preferable properties for a FT\slash{}FI benchmarking suite.

\textit{A. Different Granularities.}
We observe that the FT\slash{}FI literature uses both simple, bare-metal
algorithm implementations and complex system compositions for demonstration
purposes. Therefore we suggest to include benchmarks in different granularities,
e.g.\ 1)~isolated algorithm implementations and program parts for a more
targeted analysis and 2)~integrated systems that demonstrate more realistic use
cases.

\textit{B. Selection of Relevant Benchmarks.} We suggest that it should be
possible to classify benchmarks into different groups that can be utilized by
FT\slash FI researchers to achieve a representative and preferably minimal
experiment setup. Currently we have two categorizations in mind which are 1)
categorizing benchmarks with respect to their  program characteristics, which
might include memory usage, runtime, fault space characteristics, etc., with
which redundant benchmarks could be avoided and 2) categorizing benchmarks into
specific domains, like possibly \textit{MiBench} \cite{guthaus:2001:mibench} in
the embedded systems domain.

\textit{C. Resource-Efficient Fault Injection.} An FI benchmarking
suite should be designed with heavy FI evaluations in mind. In this regard, we
aim for a benchmark suite that 1)~relies on a lightweight infrastructure and 2)~is
\emph{configurable} to meet requirements such as execution time and memory usage.
These properties would be helpful to adjust the fault-space size to specific
needs, e.g.\ to reduce the number of necessary injections, enabling
large-scale FI experiment campaigns.

\textit{D. Self-contained Runtime.}
Besides being lightweight, we propose that the benchmarking suite should also be
self-contained and portable. On one hand, FI frameworks such as
FAIL* \cite{schirmeier:15:edcc} run the benchmarks in an instrumented virtual
machine. This requires target programs to bring their own runtime, including
a barebone operating system and in most cases, a standard library. On the other
hand, the dependency on uncertain external runtime support also makes experiment
results less comparable across different studies.
\textit{TACLeBench}~\cite{TACLeBench} shows a good example of portability in
this regard.

\section{Evaluation of Selected Properties}
\label{sec:evaluation}
In this section, we exemplarily explore some of the aforementioned benchmarking suite properties
in more detail. For this fast abstract, we try to find criteria to classify
benchmarks based on their characteristics and demonstrate what benefit a
lightweight infrastructure could bring to speed up FI.

\begin{figure}[]
	\centerline{\includegraphics[width=\columnwidth]{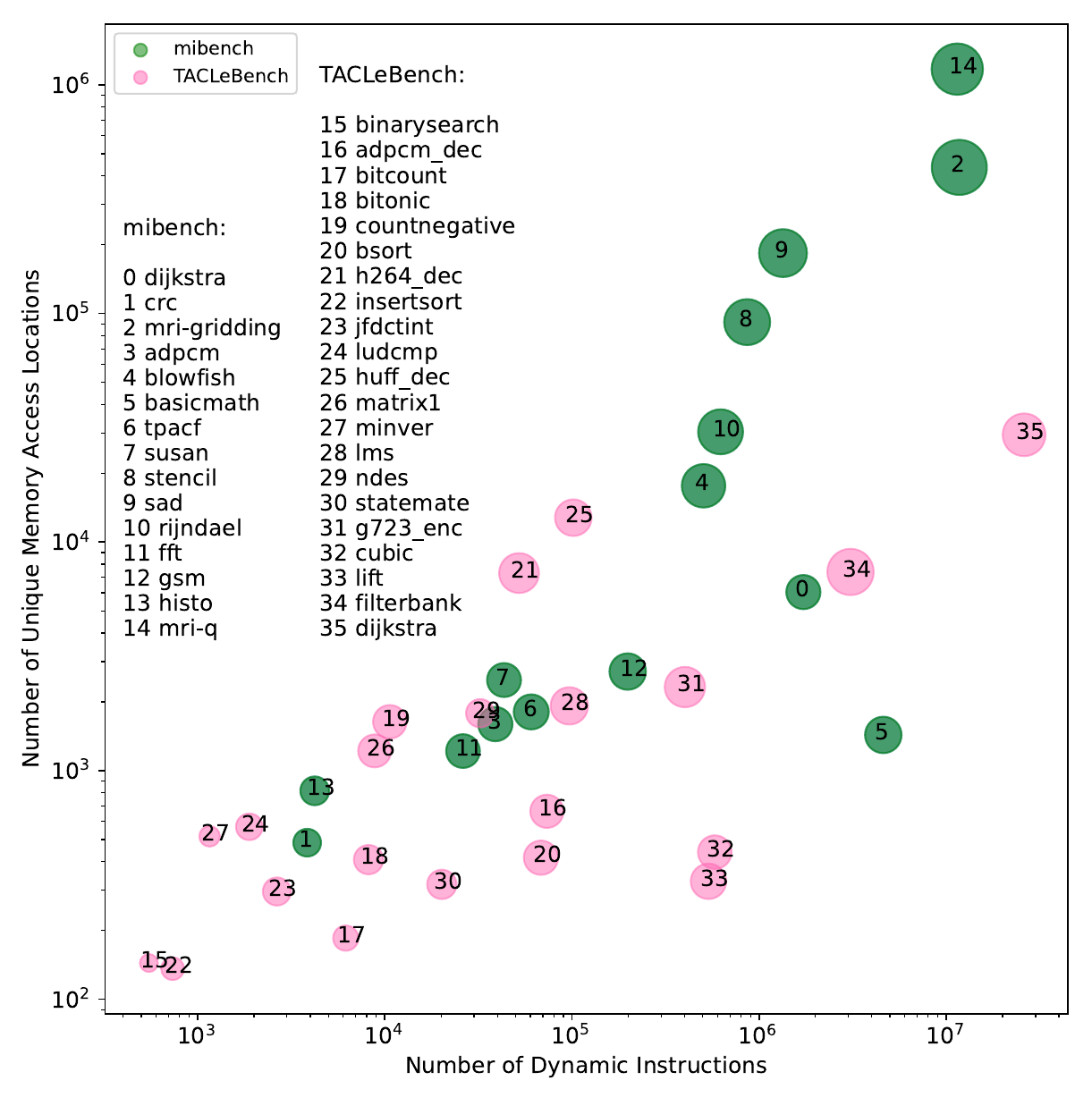}}
	\caption{
        Number of dynamic instructions, unique memory-access locations and
        SDCs (circle sizes) of \textit{MiBench} \cite{guthaus:2001:mibench}
        benchmarks compiled with \textit{picolibc} \cite{repo:picolibc:1.8.6} and \textit{TACLeBench}
        \cite{TACLeBench} as used by Borchert et al.~\cite{borchert:23:dsn}, measured with the
        \textit{FAIL*} fault-injection framework.
	}
    \label{fig:benchmark-properties}
\end{figure}

The plot in \prettyref{fig:benchmark-properties} shows some of the aforementioned properties derived from an FI campaign with 
\textit{MiBench} \cite{guthaus:2001:mibench} and \textit{TACLeBench}
\cite{TACLeBench,borchert:23:dsn}. As a preliminary study, it includes the number of dynamic
instructions, the number of memory-access locations\footnote{The number of
unique memory locations where at least one memory read or write event is
recorded during the FI campaign.},
and the \textit{Silent Data Corruption} (SDC) count.
From a bird's eye view, the benchmarks fill in three quadrants, and exhibit some clustering
patterns. This may suggest that 1)~we are missing benchmarks that have very high
data throughput (e.g., using SIMD instructions),
2)~some benchmarks may be overlapping in terms of their fault-space
characteristics, and 3)~the outliers may be interesting targets.

Interesting program characteristics beyond those shown are, e.g., stack and heap
usage, branching behavior, or memory-access granularity.

To highlight the relevance of program granularities, we measured the runtimes of a subset of \textit{MiBench} benchmarks
compiled (1)~in a full-system setting with the \textit{eCos}~\cite{massa:02:ecos-book} OS, and (2)~in a bare-metal setting based on \textit{picolibc}~\cite{repo:picolibc:1.8.6}.
The results show that
for the \textit{MiBench} benchmarks
\textit{dijkstra},
\textit{susan},
\textit{crc} and
\textit{cutcp},
the \textit{eCos} variants have 37\%--128\% more dynamic
instructions. The number of dynamic memory accesses is increased by the same order of
magnitude.
In the same experiments, the FI results are also vastly different across the two variants: Notably the
full-system setting is more prone to failure modes such as timeout. For example,
the \textit{eCos} variant of \textit{crc} has 97\% more SDCs, 801\% more
timeouts and 1491\% more CPU exceptions compared to the \textit{picolibc} variant.

\section{Conclusion}

To start a discussion on a potential dedicated FT\slash FI-specific benchmarking suite, we demonstrated how this domain could benefit.
Such a benchmarking suite should have a full coverage in terms of both practical
applicability and program characteristics, include benchmarks in different
granularities, ship a self-contained and lean runtime, and expose rich
configurability.
Nevertheless, it still remains an open question what the most relevant
program properties for FT\slash{}FI would be.

\AtNextBibliography{\footnotesize} %
\printbibliography{}

\end{document}